\documentclass[11pt,twoside]{article}
\usepackage{asp2006}
\usepackage{psfig}
\usepackage{epsf}
\usepackage{graphics}
\usepackage{lscape}
\def\edcomment#1{\iffalse\marginpar{\raggedright\sl#1\/}\else\relax\fi}
\reversemarginpar

\begin{document}
\title{High-z radio starbursts host obscured X-ray AGN}
\author{A.M.S. Richards, R. Beswick, S.T. Garrington, T.W.B. Muxlow, H. Thrall}
\affil{Jodrell Bank Observatory, University of Manchester,
  SK11 9DL, UK}
\author{M.A. Garrett, M. Kettenis, H.J. van Langevelde}
\affil{JIVE, Postbus 2, 7990 AA Dwingeloo, The Netherlands}
\author{E. Gonzalez-Solarez, N.A. Walton}
\affil{Institute of Astronomy, Madingley Road, Cambridge CB3 0HA, UK}
\author{M.G. Allen}
\affil{CDS (UMR 7550), 11 rue de l'Universit\'{e}, 67000 Strasbourg, France}

\begin{abstract}
We use Virtual Observatory methods to investigate the 
  association between radio and X-ray emission at high
  redshifts.  Fifty-five of the 92 HDF(N) sources resolved by combining
  MERLIN+VLA data were detected by {\em
  Chandra}, of which 18 are hard enough and bright
  enough to be obscured AGN. The high-$z$ population of $\mu$Jy radio sources
  is dominated by starbursts an order of magnitude
  more active and more extended than any found at $z<1$ and at least a
  quarter of these
  simultaneously host highly X-ray-luminous obscured AGN.

\end{abstract}


\vspace{-0.5cm}
\section{Radio-based classification}

A region of 100 arcmin$^2$ around the Hubble Deep Field North
  (HDFN 10-arcmin field) was observed by the UK radio interferometer
  MERLIN and the USA VLA \cite{AMSRMuxlow05}.  92 sources brighter than 40
  $\mu$Jy were found in the combined images, which reach  a noise
  level of $<4\mu$Jy using beam sizes of $0.''2 - 2''$.  These are the only
 observations apart from the HST images which provide morphological
 information. 8.4-GHz data are also available for most of the
 field \cite{AMSRRichards98} allowing the radio spectral indices to be
 calculated. 

We employed data access, crossmatching and manipulation tools now
available via the AstroGrid Workbench and the EuroVO.  These include the Vizier and Aladin
services, TopCat and a cut-out extractor for radio images which
uses the parselTongue scripting interface developed by RadioNet.
 We compare the radio data
with the {\em Chandra} CDF(N) source list \cite{AMSRAlexander03} and other GOODS
\cite{AMSRGiavalisco04} HDFN observations to investigate the properties of
active galaxies at $z\gg1$:
 \begin{enumerate} \item Can radio emission  originate from a process different from the X-ray source in
 the same galaxy?
 If so, how much radiation has a common origin and how much is separate?
\item Are  high-redshift starbursts just analogues of local ULIRGs or are they a distinct phenomenon only seen in the early universe?
\end{enumerate}

 The presence of rest-frame FIR emission \cite{AMSRGarrett02,
 AMSRElbaz02}, the HST images \cite{AMSRGiavalisco04} and other
 non-radio data were used for supporting information but not as the
 primary classification criteria.
 We were thus able to classify the specific origins of the radio
 emission in each galaxy, which may be different from the sources of
 optical and other radiation from the same object.  Compact bright
 peaks with a flat radio spectrum are probably AGN, especially if
 radio lobes extend beyond the optical limits.  Extended emission with
 a steep non-thermal spectrum is likely to be of starburst origin and
 in some cases optical knots of starformation can be seen.  The latter
 are often associated with {\em ISO}, {\em Spitzer} or SCUBA sources
 \cite{AMSRS03, AMSRBorys04, AMSRChapman05} and with interacting or distorted galaxies in
 the {\em HST} images.  Figure 1.\ shows two typical sources.  3 out of every 4 $\mu$Jy sources
are starbursts, including those with X-ray counterparts.  Three
sources have both starburst and AGN radio characteristics. 

\begin{figure}
 \plottwo{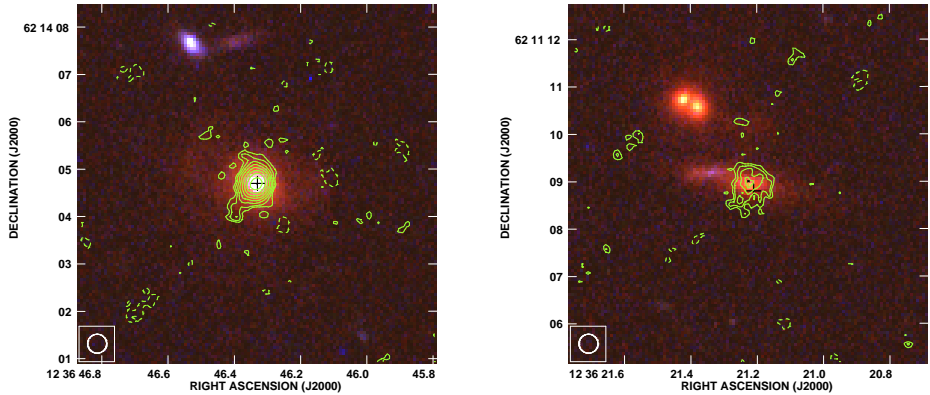}{RichardsAMS_fig2.eps}
 \caption{Radio contours overlaid on \emph{HST ACS} images.  The
   positions of the hard, obscured X-ray sources are marked with crosses.
{\itshape Left:\/} J123646+621404 ($z=0.96$) is a bright AGN with a
   compact core overlying the optical peak and lobes extending outside
   the host. It has a flat radio spectrum with $\alpha=-0.04$.
   \emph{ISO}. 
{\itshape Right:\/} 
J123621+621109 ($z=1.01$) is a radio starburst  associated with a
distorted optical system. It has a steep radio   spectrum, $\alpha>0.86$.  
}
\end{figure}

\section{Changes in the X-ray -- radio relationship at high $z$}

55 of the radio sources are among the 100 X-ray sources detected by
 {\em Chandra} in the same field \cite{AMSRAlexander03}. A {\em Chandra} source with luminosity $L_{\rm X} > 10^{37}$ W is
probably an AGN. If it has a hard photon index ($\Gamma < 1$), this
indicates the presence of an obscured (type-2) AGN \cite{AMSRAlexander05a}.
 64 type-2 AGN were identified in the 10-arcmin
field \cite{AMSRPadovani04}.  Over half
of the X-ray sources with MERLIN+VLA counterparts have $L_{\rm X} > 10^{35}$ W, 18 of which are classed as type-2 AGN. These are shown by a blue {\sf A}
in Fig.\ 2.  The radio emission is starburst dominated in at least 11 of the
 radio counterparts.   The redshift distribution of radio+X-ray source
classifications is shown in Fig. 2.

\begin{figure}
 \plotone{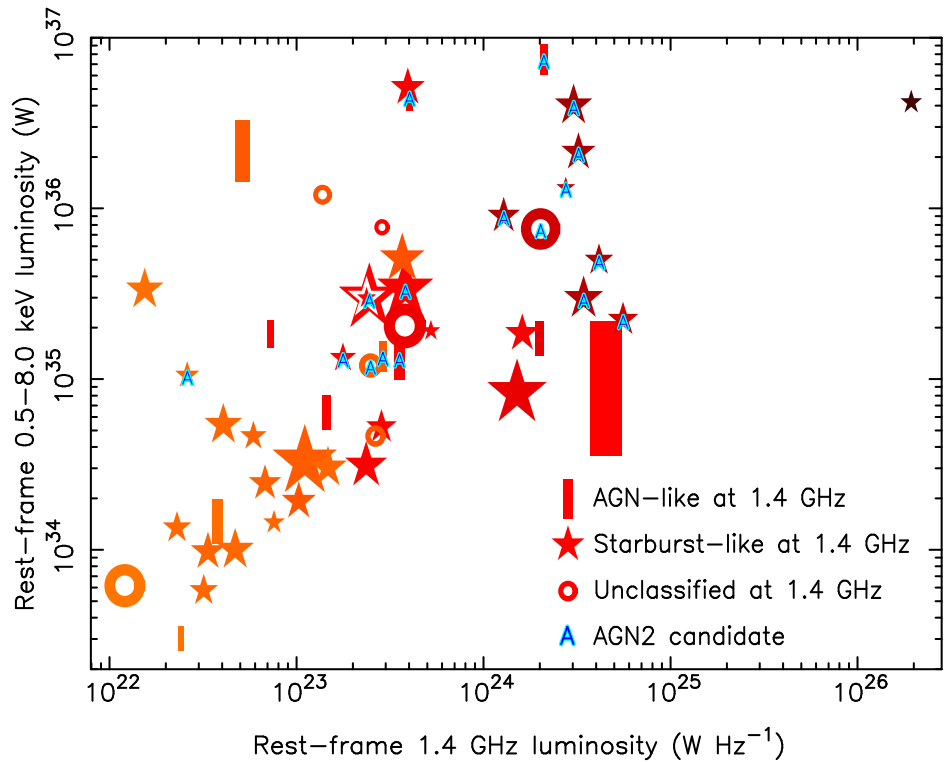}
 \caption{The distribution of the classes of radio-bright sources
           (see key) with respect to $L_{\mathrm{R}}$ and
           $L_{\mathrm{X}}$. The blue `{\bf A}'s represent X-ray
           selected AGN2.  The shade of the symbols is proportional to
           redshift, the darker sources being further away.  Broadly,
           orange symbols represent $z<1$; bright red symbols
           represent $1 < z < 2$ and browner symbols are more distant
           sources out to $z=4.424$. The size of the symbols is
           proportional to the source largest angular size. }
 \end{figure}

  At $z<1.1$, under 1/3
 X-ray sources in the HDFN have radio counterparts.  This rises to
 over 1/2 at higher redshift and in most of these the high X-ray
 luminosity indicates the presence of an AGN.  The $L_{\rm X}-L_{\rm
 R}$ relationship for starbursts at $z<1.3$ \cite{AMSRBauer02} breaks down
 at higher $z$ where we see a dramatic increase in the scatter in
 Fig. 2.  This is the case even if only the soft-band or the
 de-absorbed hard-band X-ray luminosity is considered.

Half of the 18 X-ray selected type-2 AGN with radio counterparts are at
$z>1.3$, all but one of which are radio starbursts.  Two also have
radio AGN cores.  A large proportion of these objects are SCUBA
sources, whose X-ray properties are discussed in detail by
\cite{AMSRAlexander05a}. All 4 of the radio AGN identified with X-ray
type-2 AGN, with no starburst signature, are at lower redshifts.  
The mean angular size of the radio sources in the
 HDFN is $1''.3$, corresponding to a typical extent of 8--10 kpc for
 starbursts and the inferred starformation rates are 1000--2000
 M$_{\odot}$ yr$^{-1}$ for the higher-redshift sources. This is an
 order of magnitude more exended and more vigourous than the behaviour
 of any objects at $z<1$.

\section{Conclusions}
At $z>1.3$, there is a strong association between the presence, but
  not the power, of faint radio and X-ray emission.  The extended
  radio emission is starburst-dominated in 3/4 of the objects
  \cite{AMSRMuxlow05} whilst 18 of the X-ray sources are hard
  enough and bright enough to be AGN2 \cite{AMSRPadovani04}.  11 of these
  X-ray AGN2 have radio starburst characteristics.  Their radio emission
  mechanisms must be dominated by star-formation whilst the AGN
  provides most of the X-ray luminosity. The high-$z$ population of
  $\mu$Jy radio sources contains a significant fraction of starbursts
  an order of magnitude more active and more extended than local
  ULIRGS, and they simultaneously host highly X-ray-luminous obscured
  AGN.


\end{document}